# The Modified Intrinsic Charm Model And $J/\psi$ Production in Hadronic Collisions

Yu.A.Golubkov

e–mail: golubkov@elma01.npi.msu.su,  golubkov@vxdesy.desy.de

Abstract

Using the statistical approach additional light partons have been introduced in the Intrinsic Charm model. Explicit expressions are obtained for charmed and light quark distributions. A comparison is made with standard IC model predictions. It is shown that such a modification of the IC model leads to better agreement with the $pN$ and $\pi N$ data on $J/\psi$ production at low energies and allows to avoid a contradiction to experimental data at high energy.

# 1 Preface

Evidence for an excess of charm production at high $x_F$ observed in a number of experiments on proton–proton [1] and muon–proton [2] collisions stimulated the development of the model with an assumption that the hadron wave function decomposition may contain the $|uud\bar{c}c>$ Fock state component. The first theoretical estimation for the charmed particles production due to nonpertubative charm component inside the proton has been made by S.Brodsky et al. in [3] using old–fashioned pertubation theory. The standard Intrinsic Charm (IC) model predicts very hard distributions for charmed particles due to the fact that the charmed quarks in the proton carries a large fraction of the proton longitudinal momenta. But some last experimental data [10] on the $J/\psi$ production in $pN$ interactions at $P_0 = 800\ GeV/c$ show that the distribution predicted by the model [3] strictly contradicts to the experiment at large $x_F$.

If the above assumption about excitation of the higher Fock states with heavy quarks is valid there must be much larger probability to create additional light partons. Therefore we have to consider also the Fock states with arbitrary number of light quarks and gluons. With natural restriction that this parton system as whole has to carry the hadron quantum numbers and sum of the parton momenta is to be equal to the hadron momentum. In this case the additional light partons restrict the available phase space and lead to effective decrease of the average longitudinal momentum of heavy quarks. Despite the recreated interest to IC problem no attempts have been made so far to consider this effect. This step has obvious physical basis and, as we show, leads to visible experimental consequences due to softer momentum distributions of the partons. The increase of the number of light quarks leads also to larger absolute value of total electric charge of the partonic system and to the larger cross section in the electroproduction processes.

In present paper we exploit the statistical approach [7], derive expressions for distributions of heavy and light partons within framework of the non–covariant pertubation theory and compare the $x_F$ distributions of the $J/\psi$ particle to experimental data in $\pi N$ and $pN$ collisions.

We shall speak mostly about the proton. For meson all the calculations are completely similar. The only difference is that in the meson case the lowest Fock state contains two light quarks instead of three ones and, therefore, number of integrations in the following expressions will be lesser by unit in comparison to the proton case.

# 2 Description of the model

## 2.1 The original IC model

Further in present paper we shall denote as $F^{(n)}$ a Fock state $|uudc\bar{c}+n>$ containing one $(c\bar{c})$ pair, three (two for meson) "valence" quarks, carring the hadron quantum numbers and n light partons, gluons and quark–antiquark pairs, with zero summary quantum numbers. In papers [3] it was proposed that the probability to create a Fock–state $F^{(n)}$ in the proton is being described by the expression:

$$dW^{(n)}(x_1, x_2, \ldots, x_{n+5}) \sim \left(M^2 - \sum_{i=1}^{n+5} \frac{m_{\perp i}^2}{x_i}\right)^{-2} \delta\left(1 - \sum_{i=1}^{n+5} x_i\right) \prod_{i=1}^{n+5} dx_i, \qquad (1)$$

where the $m_{\perp i}$ are the effective transverse masses of the partons $m_{\perp i}^2 = m_i^2 + <p_{\perp i}^2>$, the $<p_\perp^2>$ is the average transverse momentum squared, M is the proton mass and n denotes the number of additional light partons.

If to regard valence quarks only, i.e., the Fock state $F^{(0)} = |uudc\bar{c}>$ and to neglect mass of the proton $M$ and all the $m_\perp^2$ for light quarks one easily obtains the one–particle distributions, normalized on unit:

$$\begin{aligned} c_1^{(0)}(x) &= 1800\, x^2 \left[\tfrac{1}{3}(1-x)(1+10\,x+x^2)+2x\,(1-x)\,\ln x\right] \\ q_1^{(0)}(x) &= 6\,(1-x)^5 \end{aligned} \quad (2)$$

for charmed and light quarks, respectively.

Because we are interested only in the hadron Fock states containing two charmed quarks we can put in the folowing calculations the overall normalization $N_{IC}$ equal to unit for the sake of simplicity. To recalculate final results it is enough to multiply them by factor $N_{IC}^{exp}$, extracted either from an experiment or from a theoretical model. This normalization is order of $N_{IC}^{exp} \sim (0.5-1)\%$ (see [2, 4, 5, 6]).

## 2.2 Arbitrary number of light partons

In this section we use the statistical approach [7] to introduce arbitrary number of additional light partons. Remind that we consider all the partons are being on the mass shell and use the non–covariant pertubation theory [3]. It is also necessary to keep in mind that the results obtained in present paper are, generally speaking, valid only for partons with sufficiently large momentum fractions ("valence" partons). The wee partons are being created due to bremsstrahlung and have different distribution. In present paper we don't consider their contribution and we use the condition $\sum_i \int dx\, f_i(x) = 1$. If we introduce additional sea partons following Eq.(1) and negelect transverse masses of light partons all the light quarks and gluons will have the same momentum distribution. Thus we can not distinguish the "valence" and the "sea" light quarks. The validity of the approximation (1) for many partons we consider later in this paper.

The probability to create a final state is proportional to $d^3p/(2\pi\hbar)^3$ — the number of cells in the momentum space. Because we don't consider transverse movement we can use the longitudinal phase space only. We define $p_k = P_{max}\, x_k$, where $P_{max}$ is the maximal longitudinal momentum of the parton. Thus we can write for the probability to create either one gluon $(g)$ or one quark $(q)$ (remind, that $\hbar = 1$):

$$dW \sim \frac{b}{2\,\pi}\, w_{g,q}\, P_{max}\, dx_g \equiv a_{g,q}\, dx_{g,q}, \quad (3)$$

where, $w_k$ is the statistical weight of the parton depending on its colour and spin. To simplify further formulae we introduced the entity $a_{g,q} = (b/2\pi)\, w_{g,q}\, P_{max}$.

For the on–mass–shell gluons we have eight colours and two polarization states. Similarly for each quark we have three colours and two polarization states. Thus we have:

$$w_g = 16, \qquad w_q = 6. \tag{4}$$

If we neglect any dynamical effects we can put $b_g \approx b_q = b \approx const$. The parameter $b$ is unknown and we at present have no indication on its exact value. But we can expect that it is order of unit. This parameter has to be adjusted either from experiment or from a theoretical model. E.g., in the bag model we are able, in principle, to calculate the necessary probabilities and distributions but it is completely unclear how will these values be transformed under the Lorentz boost to the infinite momentum system. Therefore the most direct way is a comparison to experimental data.

We don't distingush the flavours of the $(q\bar{q})$ pairs, i.e., we consider the $SU(3)$–symmetric sea. This suggestion does not play any rôle in our case. So we can write the expression for the probability to create a state $F^{(n)}$ with one $(c\bar{c})$ pair, three quarks $uud$ (two quarks for a meson), $n_g$ gluons and $n_q$ quark–antiquark pairs:

$$dW_p^{(n)}(x_1, \ldots, x_{n+5}) \sim \sum_{n_g+2n_q=n} \frac{a_g^{n_g}}{n_g!} \frac{a_q^{2n_q}}{(2n_q)!} \frac{x_1^2 x_2^2}{(x_1+x_2)^2} \prod_{i=1}^{n+5} dx_i\, \delta(1 - \sum_{j=0}^{n+5} x_j). \tag{5}$$

Here $n = n_g + 2n_q$, indecies "1" and "2" denote the $c$ and $\bar{c}$ quarks, respectively. The factors $1/n_g!$ and $1/(2n_q)!$ take into account indistingushibility of the gluons and the quarks.

To obtain the total probability for all the Fock states containing two charmed quarks one needs to perform summation over $n$, $0 < n < \infty$, and the integration over all $x_i$:

$$1 = Z_p^{-1}(a_g, a_q) \sum_{n=0}^{\infty} \int dW_p^{(n)}(x_1, \ldots, x_{n+5}), \tag{6}$$

where, $Z_p^{-1}(a_g, a_q)$ is the normalization factor (we put the overall normalization equal to unit). If we perform the integration over all $x_i$ in Eq.(5) we obtain:

$$W_p^{(n)} = \int dW_p^{(n)}(x_1, \ldots, x_{n+5}) = \frac{1}{5} \frac{1}{(n+6)!} \sum_{n_g+2n_q=n} \frac{a_g^{n_g}}{n_g!} \frac{a_q^{2n_q}}{(2n_q)!} \tag{7}$$

For the normalization factor we have:

$$Z_p(a_g, a_q) = \sum_{n=0}^{\infty} W_p^{(n)} = \frac{1}{5} \sum_{n=0}^{\infty} \frac{1}{(n+6)!} \sum_{n_g+2n_q=n} \frac{a_g^{n_g}}{n_g!} \frac{a_q^{2n_q}}{(2n_q)!} \tag{8}$$

The summation over $n = n_g + 2n_q$ can be carried out using properties of the binomial sums. As result we obtain for the proton:

$$Z_p(a_g, a_q) = \frac{1}{10} \sum_{n=0}^{\infty} \frac{a_+^n + a_-^n}{n!(n+6)!} \equiv \frac{1}{5} S_6(a_+, a_-); \qquad a_\pm = a_g \pm a_q. \tag{9}$$

Similarly for meson we have

$$Z_M(a_g, a_q) \;=\; \frac{1}{10} \sum_{n=0}^{\infty} \frac{a_+^n + a_-^n}{n!(n+5)!} \;\equiv\; \frac{1}{5} S_5(a_+, a_-), \qquad (10)$$

where we denoted

$$S_m(a_g, a_q) \;=\; \frac{1}{2} \sum_{n=0}^{\infty} \frac{a_+^n + a_-^n}{n!(n+m)!}$$

$$= \; \frac{1}{2} \left\{ a_+^{-m/2} I_m(2\sqrt{a_+}) + a_-^{-m/2} I_m(2\sqrt{a_-}) \right\}.$$

Here, $I_m(z)$ is the modified Bessel function of the m–th order.

In the considered model in difference to usual parton picture the distributions of the additional light partons does not look like bremsstrahlung spectrum and its value is finite at $x \to 0$. It follows from Eq.(9) that the coefficients in the sum have a maximum at some number $n = n_{eff}$. Thus we can speak about effective or average number of light partons in the hadron. Fig.1 presents the average number and average momentum fraction of light partons versus the $P_{max}$ in the proton at the above choice of the relative weights $w_g$, $w_q$.

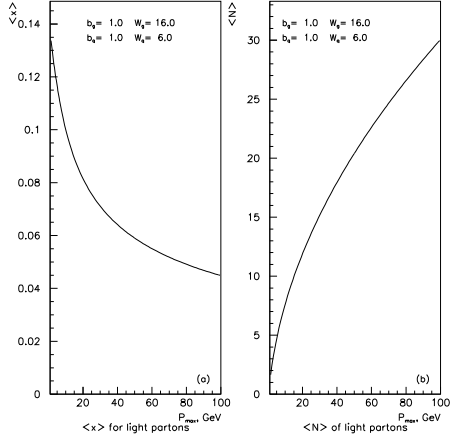

Figure 1: The average momentum fraction (a) and the average number (b) of light partons versus maximal momentum $P_{max}$ for the proton.

For light parton we obtain the following 1–particle distribution (for the proton case):

$$q_1(x; a) \;=\; \frac{1}{5} Z_p^{-1}(a_g, a_q) \, (1-x)^5 \, S_5 \left[ a_+(1-x), a_-(1-x) \right]. \qquad (11)$$

And for charmed quarks we have the 2–particle distribution:

$$c_2(x_c, x_{\bar c}; a_g, a_q) \;=\; Z_p^{-1}(a_g, a_q) \frac{x_c^2 x_{\bar c}^2}{(x_c + x_{\bar c})^2} (1 - x_c - x_{\bar c})^2 \, S_2 \left[ a_+(1 - x_c - x_{\bar c}), a_-(1 - x_c - x_{\bar c}) \right], \quad (12)$$

One–particle distribution for charmed quark looks as

$$c_1(x_c; a_g, a_q) = Z^{-1}(a_g, a_q) x_c^2 (1-x_c)^5 \int_0^1 \frac{dy\, y^2 (1-y)^2}{[x_c + (1-x_c)y]^2} S_2\left[a_+(1-x_c)(1-y), a_-(1-x_c)(1-y)\right]. \tag{13}$$

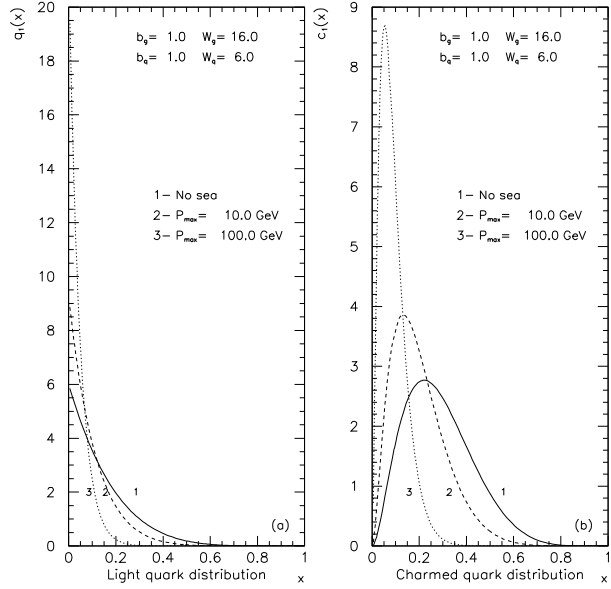

Figure 2: Distributions for light (a) and charmed (b) quarks over $x$ in the proton from Eqs.(11), (13) at $P_{max} = 10, and\, 100\ GeV$. The solid line is the distributions predicted by the IC model without additional sea partons.

Additional light partons carry zero quantum numbers and its number $n$ is not invariant, i.e., depends on the reference frame. From the definition (3) and Fig.1 we see that the average number of the light partons depends on the reaction energy. The increase in the energy of the reaction leads to the increase in the effective number of sea partons and to softer longitudinal distribution.

We have to note that the growth of the $<n>$ is faster than usual logarithmic law due to exponential growth of the number of cells in the momentum space. We assumed the constant distribution over the momentum for the unconstrained probabilities to produce a light parton. In general, it is not so and the parameter "b" can depend on the parton momentum.

## 2.3 Self–consistency of the model

Now we have to make a rough check of the self–consistency of the Eq.(1). Let us consider one Fock state with fixed number $n$ of additional light partons. Perfoming integration in Eq.(1) over necessary parton momenta it is easy to obtain the 2–particle distribution for $c\bar{c}$ pair:

$$c_2^{(n)}(x_c, x_{\bar{c}}) \sim \frac{x_c^2 x_{\bar{c}}^2}{(x_c + x_{\bar{c}})^2} (1 - x_c - x_{\bar{c}})^{n+2}, \tag{14}$$

and respectively $(n+3)$ particle distribution for light partons:

$$q_{n+3}^{(n)}(x_1, x_2, \ldots, x_{n+3}) \sim \left(1 - \sum_{i=1}^{n+3} x_i\right)^3. \tag{15}$$

These two expressions lead to 1–particle distributions:

$$c_1^{(n)}(x_c) \sim x_c^2 (1 - x_c)^{n+5} \int_0^1 dy \frac{y^2(1-y)^{n+2}}{[x_c + (1-x_c) y]^2} \tag{16}$$

for charmed quarks and

$$q_1^{(n)}(x) \sim (1-x)^{n+5} \tag{17}$$

for light quarks.

Average fractions of the proton momentum carried by a quark is:

$$\begin{aligned} <x_c> &= \tfrac{2}{n+7}, \quad \text{for c quark} \\ <x_q> &= \tfrac{1}{n+7}, \quad \text{for light parton.} \end{aligned} \tag{18}$$

All the above formulae have been derived in suggestion that effective transverse masses of heavy quarks are much larger than for light ones. It is really so but for validity of the approach we need to satisfy more restrictive condition for each Fock state $F^{(n)}$:

$$\frac{2 <m_{\perp c}^2>}{<x_c>} \gg \left| M^2 - \sum_{i=1}^{n+3} \frac{<m_{\perp i}^2>}{<x_i>} \right|$$

We take:
$$<m_{\perp c}^2> \approx 2m_c^2 \approx 4.5 \, GeV^2; \quad <x_c> \approx 2 <x_q>$$

For $<m_{\perp q}^2> \approx <p_{\perp q}^2> \approx 0.1 \, GeV^2$ and for $<x_q> \approx 1/(n+7)$ one can obtain that for $n < 12$ the above approximation is valid within 30% accuracy. We hope that general formulae from previous subsection presenting the sum over all states work due to double factorial in the denominators of the expansion coefficients, which suppresses the contributions of the higher states. Of course, at small $x_q$ where the distribution $q_1(x) \gg c_1(x)$ the validity of the approximation will be definetely brocken but this model considers large values of x.

So we conclude that our expressions are self–consistent for sufficiently large $P_{max} \leq 20$ (see Fig.1(b)).

# 3   Application to $J/\psi$ Production

## 3.1   The recombination model

To estimate the longitudinal distributions of the $J/\psi$ particles we use the recombination model [8]. In this model the differential cross section of final particle can be written as:

$$\frac{d\sigma}{dx_F} \sim \int_0^x dx_1\, dx_2\, F(x_1, x_2)\, R(x_1, x_2; x_F), \qquad (19)$$

where, $F(x_1, x_2)$ is the two–particle distribution for quarks "1" and "2" with momentum fractions $x_1$ and $x_2$, respectively; $R(x_1, x_2; x_F)$ is the recombination function describing the probability for two quarks to coalesce in the final meson with momentum fraction $x_F$. In the simplest case:

$$R(x_1, x_2; x_F) = \delta(x_F - x_1 - x_2),$$

ensuring the longitudinal momentum conservation. In principle for charmed particles the primordial transverse momenta of the initial $c$ quarks can reach large values ($\geq 1\ GeV$) and have to be taken into account by obvious way (see, e.g., [9]). But for the aim of present paper it is not important. As a result from Eqs.(19) and (12) one obtains for the $x_F$ distribution of $J/\psi$ particles the following expressions:

$$\begin{aligned}
\frac{d\sigma(J/\psi)(pp)}{dx_F} &= \tfrac{1}{30} Z_p^{-1}(a_g, a_q)\, x_F^3\, (1 - x_F)^2\, S_2\left[a_+(1 - x_F), a_-(1 - x_F)\right], \\[6pt]
\frac{d\sigma(J/\psi)(\pi p)}{dx_F} &= \tfrac{1}{30} Z_M^{-1}(a_g, a_q)\, x_F^3\, (1 - x_F)\, S_2\left[a_+(1 - x_F), a_-(1 - x_F)\right],
\end{aligned} \qquad (20)$$

for $pp$ and $\pi p$ collisions, respectively. For hadronic interactions we have in the center–of–mass system $P_{max} = \sqrt{s}/2$. If we normalize our cross section on specific value we have the only free parameter — the constant $b$ from Eq.(3) At the moment we have no other choice as to put it equal to unit, $b = 1$. Note here that at larger values of $b$ we have softer distributions for $c$ quarks and for $J/\psi$ particle.

In paper [10] the $J/\psi$ production in the proton–nucleus interactions has been measured at $\sqrt{s} = 38.7\ GeV$ and $0.30 < x_F < 0.95$. It is a region where we expect the contribution from the intrinsic charm. Based on estimations from [11] and private communication to R.Vogt the authors of the paper [10] made a conclusion that the IC model conradicts to their experimental data. They used a value for total cross section for IC $J/\psi$ production $\sigma_{tot}(J/\psi) \approx 1.8\ nb$. The present paper is not right place to discuss the absolute value of the $J/\psi$ total cross section. It is very complicated subject and needs special consideration. So we use the value $0.9\ nb$ for $\sigma(J/\psi)$ in the forward hemisphere at FNAL energy. For lower energy data [12] we normalized our expressions for measured total cross section in the forward direction for the so called "diffractive" component (see [12]) of the $J/\psi$ yield.

Results of this comparison are shown in Figs.3 and 4 for $b = 1$. In these figures we also plotted the predictions of the standard IC model based on the same recombination model and with the same normalization. As one observes the predictions of the usual IC model can really be rejected for above value of the total cross section at FNAL energy. On another hand predictions followed from our calculations don't contradict to the experimental data. At low energies both models agrees with the experimental data.

This fact allows us to suggest that the contribution from intrinsic charm can be observed at sufficiently low initial energies and is practically negligible at high enrgy even at large $x_F$. This comparison also confirms that parameter $b \sim 1$.

## Acknowledgement

I want to thank Dr. G.Wolf very much for useful discussions. I would like to express my gratitude to DESY for partial support during this work.

I am also thankful to Prof. P.F.Ermolov for his stimulating interest to this subject.

# References

[1] *D.Drijard et al.*, Phys. Lett. **81B** (1979) 250; **85B** (1979) 452; *K.L.Giboni et al.*, Phys. Lett. **85B** (1979) 437; *W.Lockman et al.*, Phys. Lett. **85B** (1979) 443; *A.Chilingarov et al.*, Phys. Lett. **83B** (1979) 136.

[2] *European Muon Collaboration. J.J.Aubert et al.* Phys. Lett. **110B** (1982) 73; Nucl. Phys. **B213** (1983) 31.

[3] *S.J.Brodsky, P.Hoyer, C.Peterson and N.Sakai.* Phys. Lett., **93B** (1980) 451; *S.J.Brodsky and C.Peterson.* Phys.Rev. **D23** (1981) 2745.

[4] *E.Hoffmann and R.Moore.* Z.Phys. **C20** (1983) 71.

[5] *Yu.A.Golubkov.* Preprint DESY 94–060, 1994.

[6] *B.W.Harris, J.Smith and R.Vogt.* Nucl. Phys. **B461** (1996) 181.

[7] *J.Kuti and V.F.Weisskopf.* Phys. Rev. **D4** (1971) 3418.

[8] *K.P.Das and R.C.Hwa.* Phys. Lett. **68B** (1977) 459.

[9] *Yu.A.Golubkov, R.V.Konoplich and Yu.P.Nikitin.* Preprint Rome–245–1981, Mar. 1981; Sov. J. Nucl. Phys, **35** (1982) 239.

[10] *M.S.Kowit et al.* Phys. Rev. Lett. **72** (1994) 1318.

[11] *R.Vogt, S.J.Brodsky and P.Hoyer.* Nucl.Phys., **B360** (1991) 67; *S.J.Brodsky and P.Hoyer.* Phys. Rev. Lett., **63** (1989) 1566.

[12] *J.Badier et al.* Z.Phys. **C20** (1983) 101.

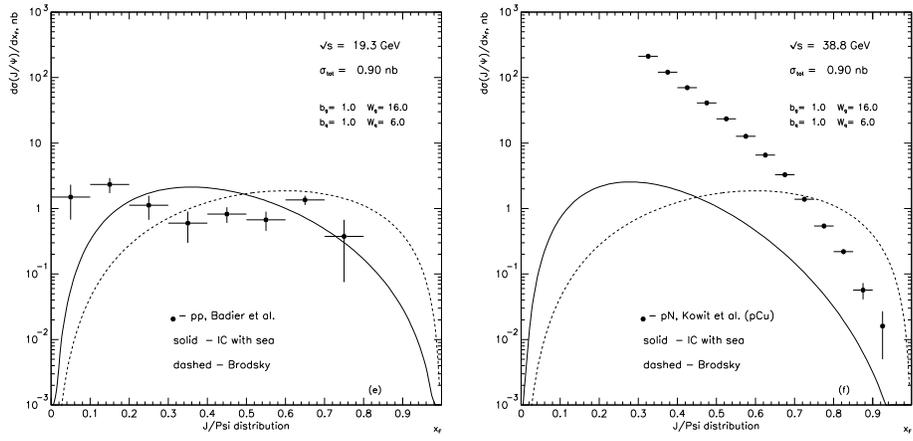

Figure 3: $J/\psi$ distributions over $x_F$ in (a) $pp$ at $P_0 = 200\ GeV/c$ and (b) in $pCu$ collisions at $P_0 = 800\ GeV/c$.

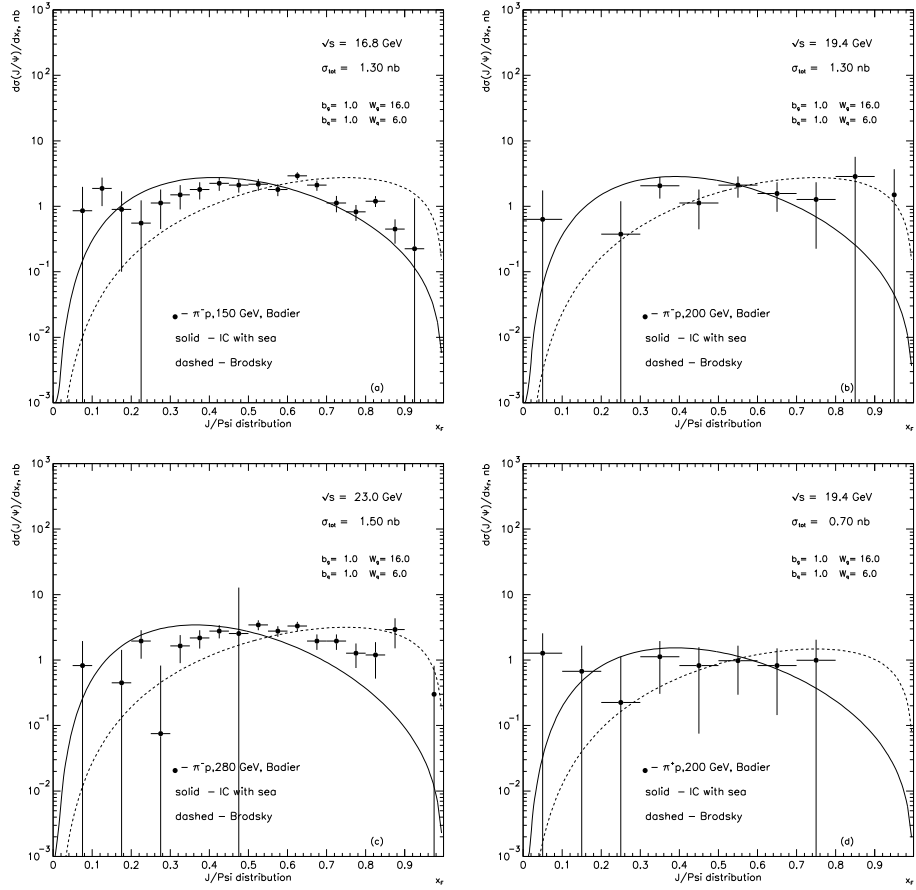

Figure 4: $J/\psi$ distributions over $x_F$ in $\pi^- p$ collisions at (a) $P_0 = 150\ GeV/c$; (b) $P_0 = 200\ GeV/c$; (c) $P_0 = 280\ GeV/c$ and (d) in $\pi^+ p$ collisions at $P_0 = 200\ GeV/c$.